\documentclass{article}

\usepackage{PRIMEarxiv}

\usepackage[utf8]{inputenc} 
\usepackage[T1]{fontenc}    
\usepackage{url}            
\usepackage{booktabs}       
\usepackage{amsfonts}       
\usepackage{nicefrac}       
\usepackage{microtype}      
\usepackage{lipsum}
\usepackage{fancyhdr}       
\usepackage{graphicx}       
\graphicspath{{media/}}     
\usepackage{subcaption}
\usepackage{colortbl}
\usepackage{xcolor}
\usepackage{arydshln}
\usepackage{tabularx}
\usepackage[
    colorlinks=true,  
    linkcolor=black,  
    citecolor=black,  
    urlcolor=black    
]{hyperref}
\usepackage{float}

\title{Lightweight G-YOLOv11: Advancing Efficient Fracture Detection in Pediatric Wrist X-rays
}

\author{
  Abdesselam Ferdi \\
  Electronics Department \\
  Constantine 1 - Frères Mentouri University \\
  Constantine, Algeria\\
  \texttt{abdesselam.ferdi@gmail.com} \\
}

\begin{document}
\maketitle

\begin{abstract}
Computer-aided diagnosis (CAD) systems have greatly improved the interpretation of medical images by radiologists and surgeons. However, current CAD systems for fracture detection in X-ray images primarily rely on large, resource-intensive detectors, which limits their practicality in clinical settings. To address this limitation, we propose a novel lightweight CAD system based on the YOLO detector for fracture detection. This system, named ghost convolution-based YOLOv11 (G-YOLOv11), builds on the latest version of the YOLO detector family and incorporates the ghost convolution operation for feature extraction. The ghost convolution operation generates the same number of feature maps as traditional convolution but requires fewer linear operations, thereby reducing the detector's computational resource requirements. We evaluated the performance of the proposed G-YOLOv11 detector on the GRAZPEDWRI-DX dataset, achieving an mAP@0.5 of $0.535$ with an inference time of $2.4$ ms on an NVIDIA A10 GPU. Compared to the standard YOLOv11l, G-YOLOv11l achieved reductions of $13.6\%$ in mAP@0.5 and $68.7\%$ in size. These results establish a new state-of-the-art benchmark in terms of efficiency, outperforming existing detectors. Code and models are available at \href{https://github.com/AbdesselamFerdi/G-YOLOv11}{https://github.com/AbdesselamFerdi/G-YOLOv11}.
\end{abstract}


\section{Introduction}
Pediatric wrist fractures are a common medical condition in children that require timely and accurate diagnosis to prevent complications such as inadequate healing, deformities, or long-term functional impairments. Traditional computer-aided diagnosis (CAD) systems have relied on handcrafted features and rule-based algorithms to aid radiologists in interpreting X-ray images. However, with the advent of deep learning, CAD systems have undergone significant advancements, leveraging the power of convolutional neural networks (CNNs), transformers, and other advanced architectures. Deep learning models, particularly CNNs \cite{ferdiclassification, ferdi2022colorization, ferdi2022u, ferdi2024residual}, can automatically extract hierarchical features from medical images without manual intervention, making them highly effective for analyzing complex patterns such as fractures.

The you only look once (YOLO) family of CNN-based algorithms, starting with YOLOv1 in 2015 and extending to the latest version, YOLOv11 in 2024, is designed for real-time object detection. These detectors strike a balance between speed and accuracy, enabling the rapid and precise detection of objects within images.

In the context of fracture detection, existing research can be classified into one-stage and two-stage methods, depending on the techniques employed. In the two-stage category, recent work by Wang et al. \cite{wang2021parallelnet} introduced ParallelNet, a two-stage R-CNN network with a TripleNet backbone for fracture detection using a dataset of $3,842$ thigh fracture X-ray images, achieving an average precision at an intersection over union (IoU) threshold of $0.5$ (AP@0.5) of $0.878$. Joshi et al. \cite{joshi2022deep} proposed a modified mask region-based CNN for fracture detection and segmentation using two datasets: a surface crack image dataset ($3,000$ images) and a wrist fracture dataset ($315$ images). The model was trained on the surface crack dataset and fine-tuned on the wrist fracture dataset, achieving an AP@0.5 of $0.923$ for detection. Further details on two-stage fracture detection techniques are provided by Ahmed et al. \cite{ahmed2024enhancing}. In comparison, single-stage techniques, such as YOLO detectors, have demonstrated superior performance in fracture detection \cite{ahmed2024enhancing}. Recently, YOLO detectors have been applied to pediatric wrist fracture localization. Since the release of the GRAZPEDWRI-DX pediatric wrist trauma X-ray dataset by the University Hospital Graz in Austria \cite{nagy2022pediatric} in 2022, Ju et al. \cite{ju2023fracture} proposed the YOLOv8 detector for localizing pediatric wrist fractures in X-ray images, achieving a mean AP@0.5 (mAP@0.5) of $0.638$ on this dataset. Subsequent research has focused on enhancing the performance of YOLOv8, YOLOv9 \cite{chien2024yolov9}, YOLOv10 \cite{ahmed2024pediatric}, and YOLOv11 \cite{das5056626detection} detectors by integrating attention modules \cite{chien2024yolov8} and global context blocks \cite{ju2024global}. However, these studies have primarily concentrated on improving accuracy, often overlooking critical factors such as computational efficiency (quantified by the number of floating point operations per second, or FLOPs) and memory requirements (determined by the number of parameters). The complexity of these detectors poses significant challenges for their deployment in CAD systems. Consequently, developing lightweight detectors that retain high performance while minimizing computational and memory demands is of utmost importance. This raises the question: how can we effectively transform a heavyweight detector into a more efficient, lightweight counterpart?

\begin{figure}[t]
  \centering
  \includegraphics[width=\linewidth]{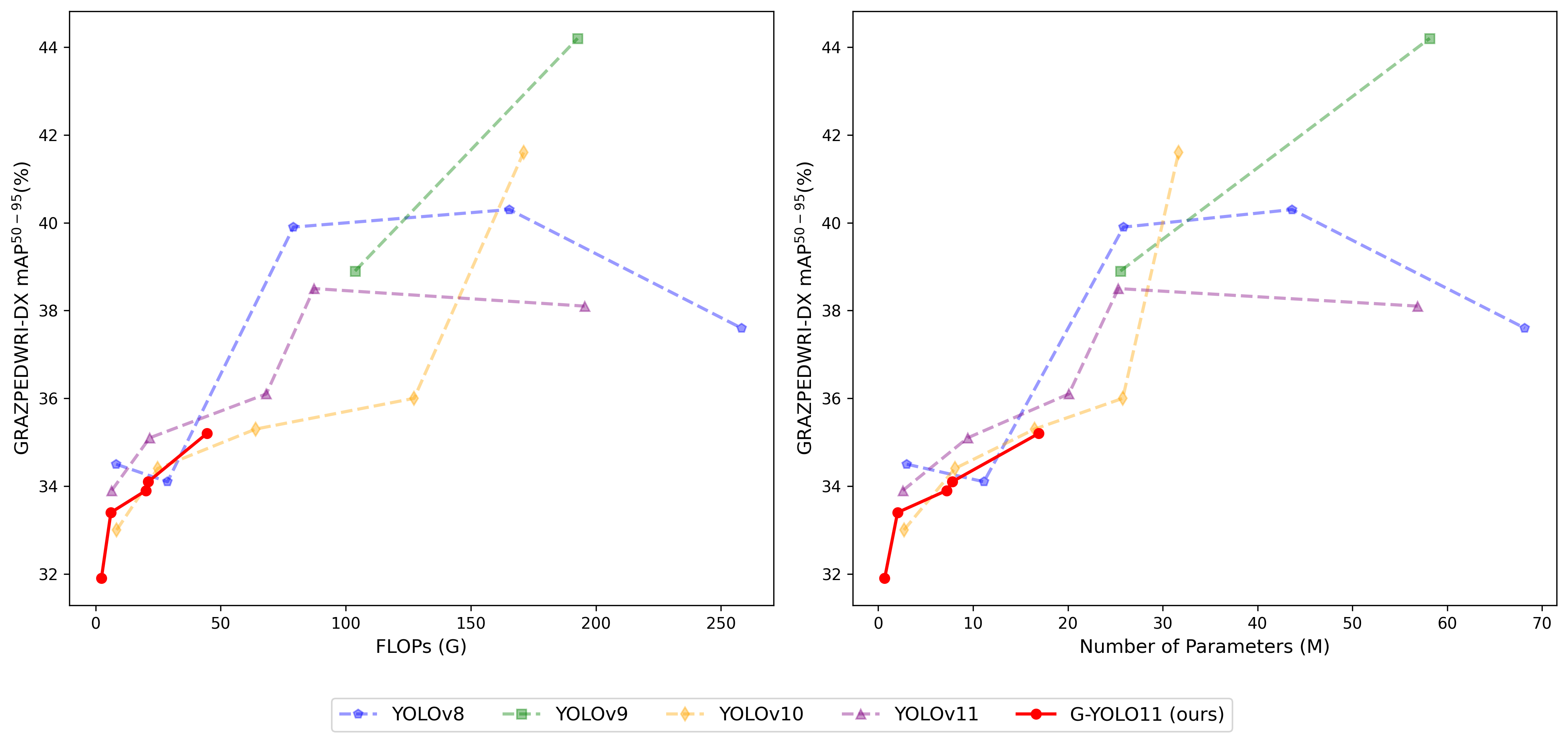}
  \caption{Comparisons of real-time object detectors on the GRAZPEDWRI-DX dataset, evaluated in terms of FLOPs-accuracy (left) and size-accuracy (right) trade-offs.}
  \label{figure1}
\end{figure}

Numerous methods have been proposed to address this challenge, with model compression being one of the most prevalent approaches \cite{ferdi2024deep, ferdi2024electrostatic}. Based on the observation that standard convolutional layers often contain redundant feature maps due to repeated information across maps \cite{han2020ghostnet}, we developed a novel lightweight YOLO detector, called ghost convolution-based YOLOv11 (G-YOLOv11), for localizing pediatric wrist trauma in X-ray images. This detector is based on the latest version of the YOLO family, YOLOv11, and incorporates the ghost convolution layer. Specifically, the standard convolution and C3k2 modules within YOLOv11 were replaced with the ghost convolution and C3ghost modules, respectively. Compared to traditional convolution, ghost convolution is computationally more efficient, generating more feature maps with fewer computationally expensive linear operations. As a result, the computational complexity of the proposed G-YOLOv11 was significantly reduced compared to standard YOLOv11 detectors, as shown in Figure \ref{figure1}. For example, our medium G-YOLOv11 (G-YOLOv11m) has $7.202$ million parameters, representing a $64.2\%$ reduction compared to YOLOv11m’s $20.095$ million parameters. Additionally, our extra-large G-YOLOv11 (G-YOLOv11x) requires $44.5$ giga FLOPs (GFLOPs), a $77.2\%$ decrease from YOLOv11x’s $195.5$ GFLOPs.

\begin{figure}[t]
  \centering
  \includegraphics[width=\linewidth]{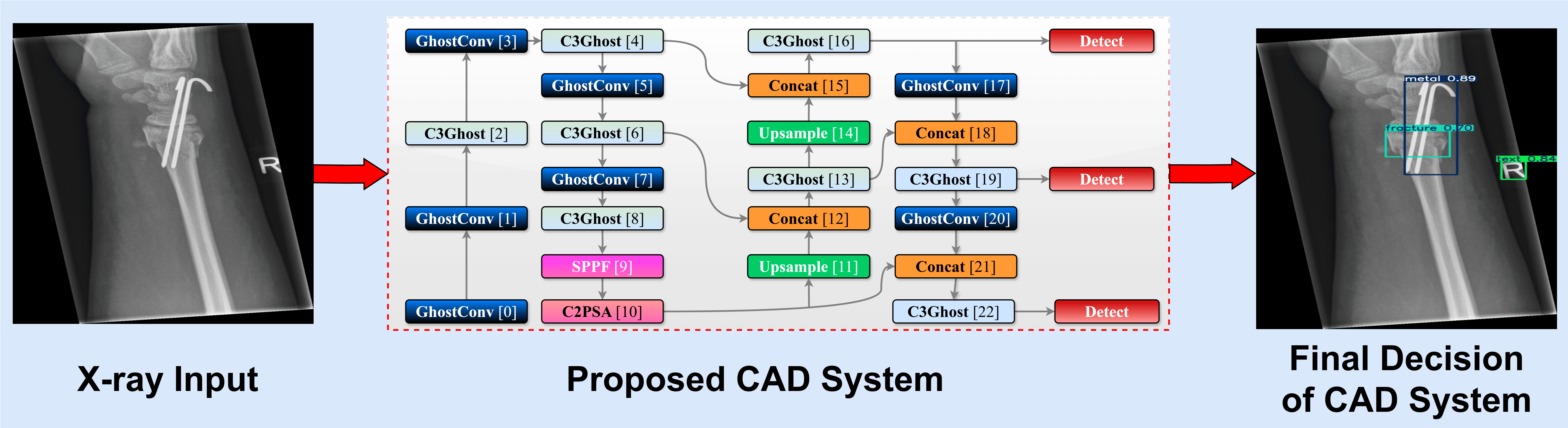}
  \caption{Schematic diagram of the proposed CAD system based on the G-YOLOv11 detector for localizing pediatric wrist trauma in X-ray images.}
  \label{figure2}
\end{figure}

Our contributions are as follows: 
\begin{itemize}
    \item We propose a novel YOLO-based CAD system (Figure \ref{figure2}) designed to assist radiologists and orthopedic surgeons in interpreting X-ray images for the detection of pediatric wrist trauma.
    \item The proposed G-YOLOv11 detector significantly reduces the computational resource requirements of the YOLOv11 detector, making it suitable for application in clinical settings.
    \item The large G-YOLOv11 (G-YOLOv11l) detector achieves state-of-the-art results in terms of efficiency, with an mAP@0.5 of $0.535$ and an inference time of $2.4$ ms on an NVIDIA A10 GPU.
\end{itemize} 

The rest of the paper is organized as follows. Section \ref{section2} describes the experimental setup and provides an overview of the proposed detector. In Section \ref{section3}, we present the obtained results, followed by a discussion of these results in Section \ref{section4}. Finally, Section \ref{section5} summarizes the work and explores potential directions for future research.

\section{Materials and Methods}
\label{section2}
This section first describes the dataset used, followed by an overview of the proposed G-YOLOv11 detector. It then presents the model’s training configuration, and concludes with a discussion of the performance evaluation metrics.

\subsection{Dataset}
The publicly available GRAZPEDWRI-DX dataset \cite{nagy2022pediatric}, which contains $20,327$ X-ray images in PNG format of pediatric wrist trauma, was used in our experiments. These images were collected from $6,091$ patients between $2008$ and $2018$ at the Department of Pediatric Surgery, University Hospital Graz, Austria. The dataset has been annotated by multiple pediatric radiologists using lines, bounding boxes, or polygons to mark pathologies such as fractures and periosteal reactions.

\begin{figure}[t]
  \centering
  \includegraphics[height=12cm, width=\linewidth]{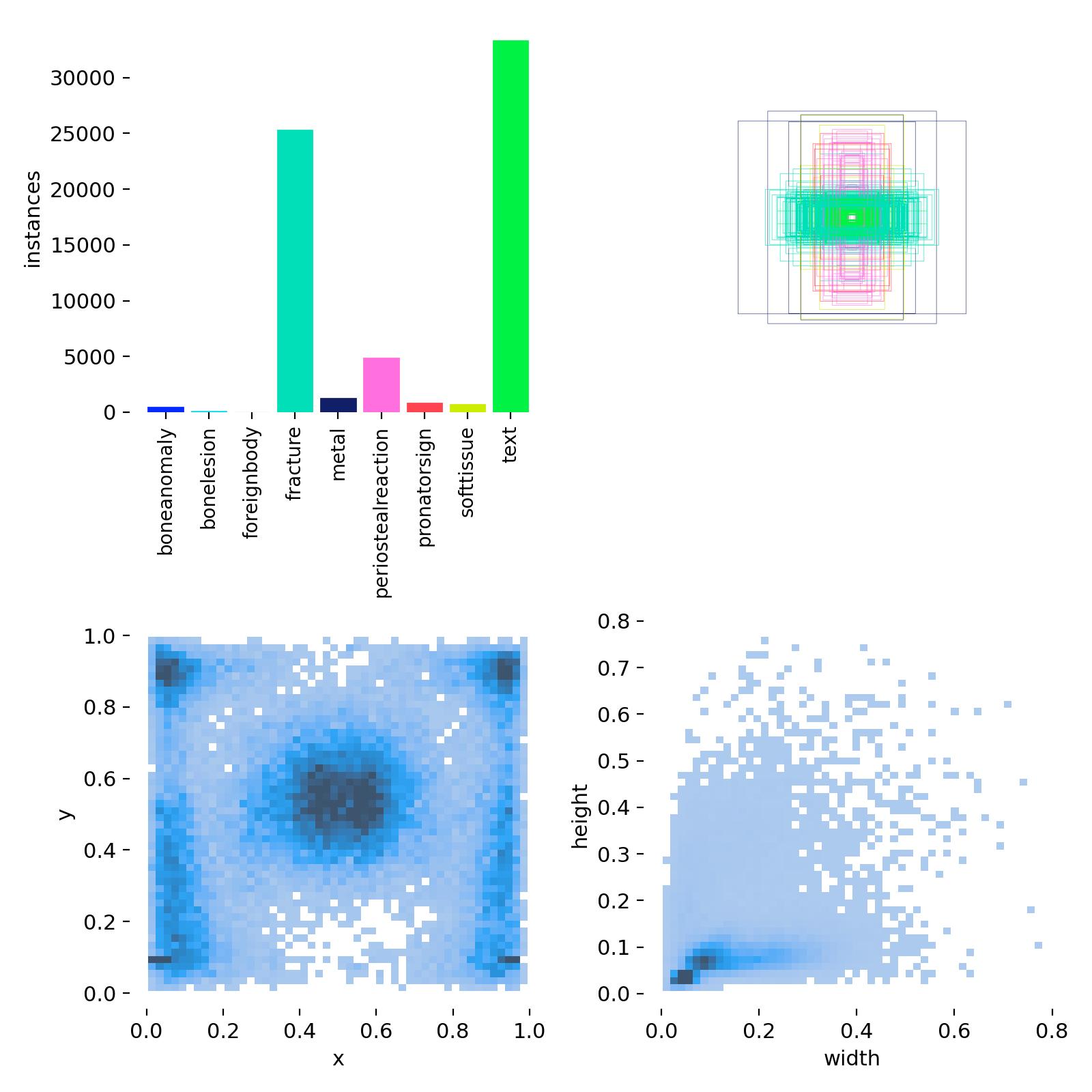}
  \caption{Visualization of training labels in the GRAZPEDWRI-DX training set. First row: distribution of classes and visualization of bounding boxes. Second row: statistical distribution of the location and size of bounding boxes.}
  \label{figure3}
\end{figure}

\begin{figure}[t]
  \centering
 \centering
\begin{subfigure}{0.4956\textwidth}
    \centering 
    \includegraphics[height=5.5cm, width=7cm]{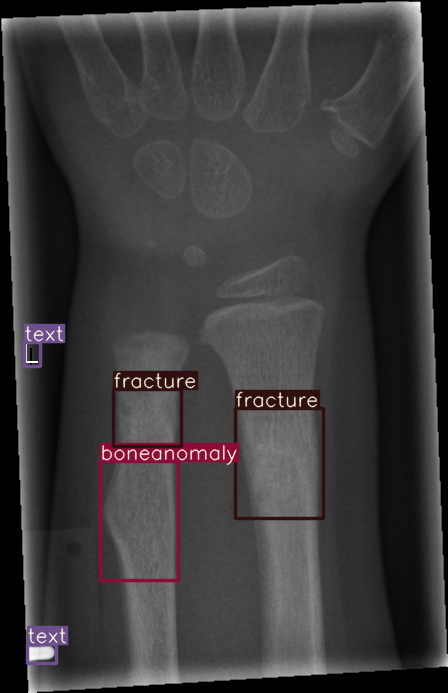}
\end{subfigure}
\hfill
\begin{subfigure}{0.4956\textwidth}
    \centering
    \includegraphics[height=5.5cm, width=7cm]{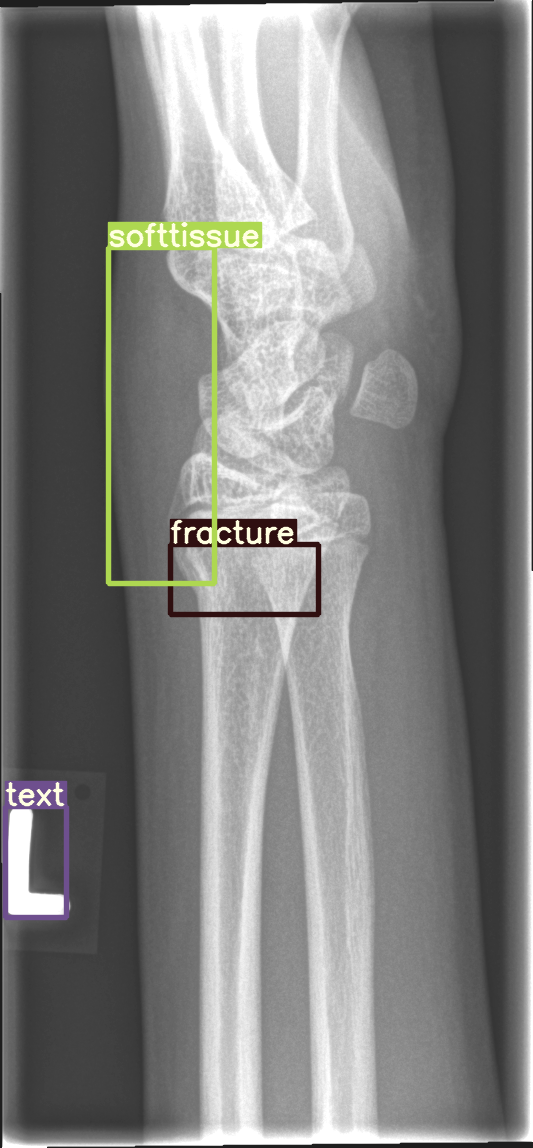}
\end{subfigure}
  \caption{Bounding box-annotated X-ray images of pediatric wrist trauma from the GRAZPEDWRI-DX dataset \cite{nagy2022pediatric}.}
  \label{figure4}
\end{figure}

For our experiments, we selected the GRAZPEDWRI-DX images with bounding box annotations in YOLO format, which contains nine classes. Following the data division and augmentation approach outlined by \cite{chien2024yolov9}, the dataset was randomly divided into $70\%$ training ($14,234$ images), $20\%$ validation ($4,065$ images), and $10\%$ testing ($2,028$ images) sets. To further augment the training data, the number of images in the training set was doubled by adjusting the contrast and luminance of the images. The label distribution and bounding box statistics are shown in Figure \ref{figure3}. Examples of two training images with bounding box annotations from the GRAZPEDWRI-DX dataset are shown in Figure \ref{figure4}.

\subsection{Proposed Detector (G-YOLOv11)}
YOLO detectors are a family of real-time object detection algorithms that have revolutionized computer vision tasks \cite{ferdi2023yolov3, ferdi2024quadratic}. Introduced in 2015 by Redmon et al. \cite{khanam2024yolov11}, YOLO detectors directly predict bounding boxes and class probabilities from input images, bypassing traditional region proposal-based methods. The series has evolved through YOLOv2, YOLOv3, YOLOv4, YOLOv5, YOLOv6, YOLOv7, YOLOv8, YOLOv9, and YOLOv10, with each iteration improving accuracy and efficiency. YOLOv11, the latest version developed in 2024 by Ultralytics \cite{yolo11_ultralytics}, offers enhanced accuracy, real-time processing capabilities, and improved efficiency, making it suitable for applications in autonomous vehicles, robotics, and medical imaging research.

YOLOv11 is available in five sizes: nano, small, medium, large, and extra-large. These sizes are determined by varying the number of convolution filters and C3k2 modules (blocks of convolutional layers for feature extraction) within the YOLOv11 architecture, as shown in Table \ref{table1}. The architecture consists of four primary components: the input, backbone, neck, and head.

\subsubsection{Input}
The input to the proposed detector consists of radiographic images, specifically X-ray images, along with the corresponding bounding box coordinates that delineate objects within these images. To enhance the detector's ability to generalize, various data augmentation techniques are applied to the input X-ray images. These include mixup (which blends two images and their labels to create a composite image) and mosaic (which combines four images into a single image).

\subsubsection{Backbone}
The backbone is responsible for extracting multi-scale features from the input images. This process involves stacking convolutional layers and specialized modules to generate feature maps of varying sizes. YOLOv11's backbone retains a structure similar to that of its predecessors, employing initial convolutional layers to downsample the input image. These layers establish the foundation for feature extraction by progressively reducing spatial dimensions while increasing the number of channels. A key advancement in YOLOv11 is the integration of the C3k2 block, which replaces the C2f block used in earlier versions. The C3k2 block provides a more computationally efficient implementation of the cross-stage partial (CSP) bottleneck by using two smaller convolutions instead of a single large convolution, as was the case in YOLOv8. Additionally, YOLOv11 retains the spatial pyramid pooling-fast (SPPF) block from previous versions but introduces a novel cross-stage partial with spatial attention (C2PSA) module.

\begin{figure}[h]
  \centering
  \includegraphics[height=12cm, width=\linewidth]{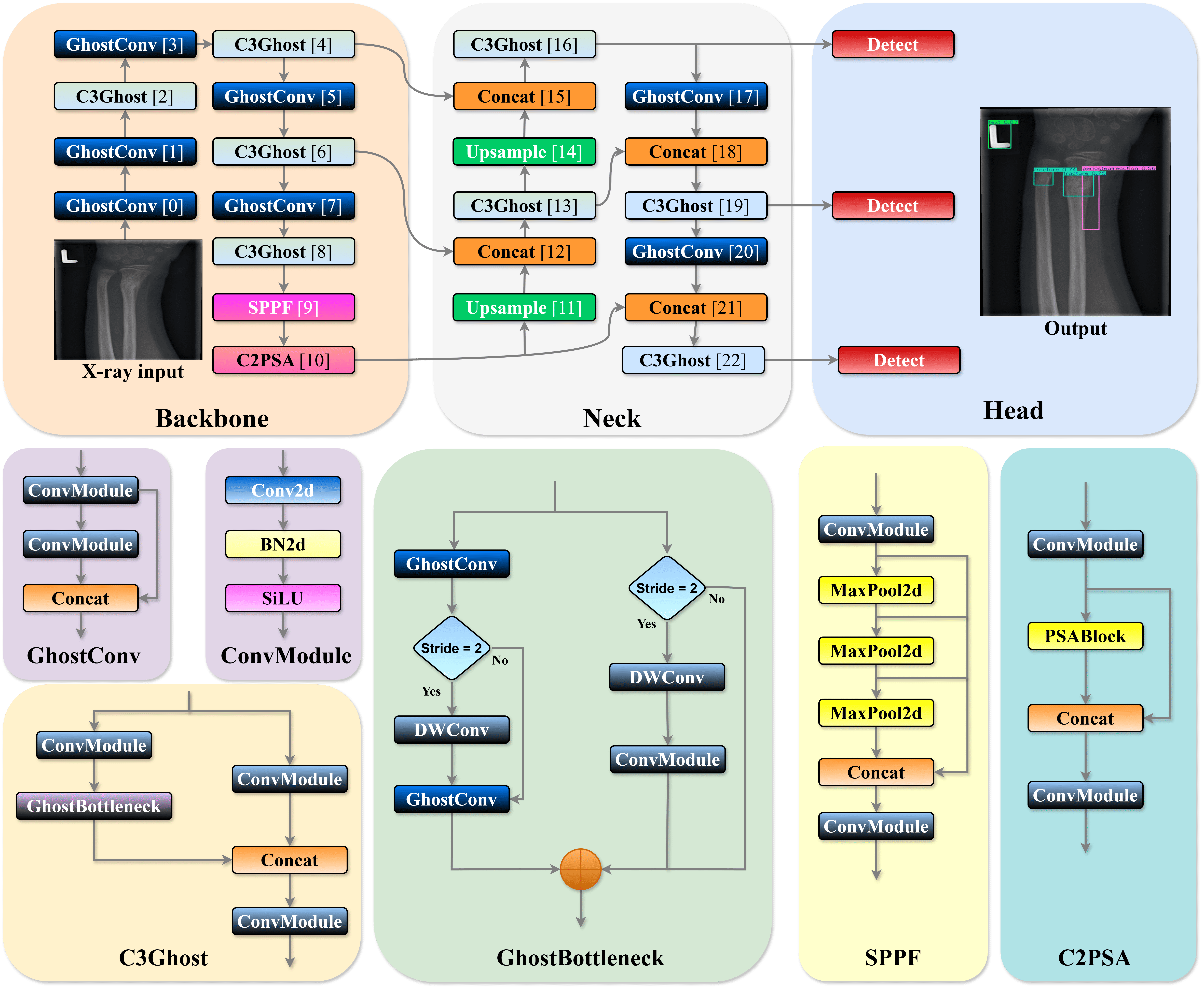}
  \caption{Architecture of the proposed G-YOLOv11 detector.}
  \label{figure5}
\end{figure}

Initially, we employed the YOLOv11 detector for the localization of pediatric wrist trauma in X-ray images. However, the complexity of this detector poses challenges for its deployment in CAD systems. Therefore, it is crucial to reduce the number of parameters and FLOPs in this detector. To address this, we developed a novel lightweight detector, named G-YOLOv11, which achieves similar detection performance with fewer computations by incorporating ghost convolution and C3Ghost modules within the YOLOv11 architecture. The architecture of the proposed G-YOLOv11 detector is shown in Figure \ref{figure5}.

\subsubsubsection{Ghost Convolution}
In standard convolutional layers, the output often contains redundant feature maps due to repeated information across the maps \cite{han2020ghostnet}. To address this inefficiency, the proposed detector incorporates the concept of ghost convolution. This new convolutional layer significantly reduces the computational complexity of the detector while maintaining the desired number of feature maps. Unlike standard convolution, ghost convolution is computationally efficient and operates consistently regardless of the input feature map size, as illustrated in Figure \ref{figure5}. The process begins with a $1 \times 1$ convolution that extracts the intrinsic feature maps. This is followed by a $5 \times 5$ convolution that generates additional feature maps using less computationally expensive linear operations. Finally, the outputs from these two layers are concatenated to produce the final feature maps, ensuring an optimized balance between efficiency and performance.

\subsubsubsection{C3Ghost}
Building on the advantages of ghost convolution, the C3k2 module in the original YOLOv11 detector was replaced with the C3Ghost module in the proposed G-YOLOv11 detector. This modification allows for the extraction of more features using efficient linear operations, significantly reducing the detector's complexity.

The C3Ghost module is designed by integrating a ghost bottleneck module with three convolutional layers, as shown in Figure \ref{figure5}. The ghost bottleneck module comprises two primary paths: the ghost convolution path and the shortcut path. The ghost convolution path includes two ghost convolution modules. The first ghost convolution module acts as an expansion layer, increasing the number of channels, while the second ghost convolution module reduces the channel count. A depth-wise convolution is positioned between these two ghost convolution modules to align with the shortcut path. The outputs from both the ghost convolution and shortcut paths are then concatenated to form the final output of the ghost bottleneck module, as depicted in Figure \ref{figure5}.

\subsubsubsection{Spatial Pyramid Pooling-Fast}
The SPPF module enables the generation of fixed-dimensional feature maps independent of input size, without the need for image resizing, which can degrade detection performance. This is achieved by applying max-pooling operations at three different levels, ensuring consistent feature extraction and preserving the detection capability of the detector.

\subsubsubsection{Cross-Stage Partial with Spatial Attention}
The C2PSA module employs convolution modules with attention mechanisms to enhance feature extraction and processing capabilities.

\subsubsection{Neck}
The neck of YOLOv11 combines multi-scale features from the backbone and transmits them to the head for prediction. It incorporates multiple C3k2 modules placed along several pathways to refine and process multi-scale features at different depths. Following the C3k2 modules, the architecture employs several convolution modules, which play a critical role in feature refinement. These layers contribute to accurate object detection by extracting relevant features, stabilizing data flow through batch normalization, and introducing non-linearity using the sigmoid linear unit activation function, which enhances model performance. Convolution modules serve as essential elements in both feature extraction and the detection process, ensuring that optimized feature maps are passed to subsequent layers.

In the G-YOLOv11 detector, the convolution and C3k2 modules in the neck were replaced with the ghost convolution and C3Ghost modules, respectively, to reduce the complexity of the neck while maintaining detection performance.

\subsubsection{Head}
The head is responsible for generating the final predictions. It processes the feature maps provided by the neck using the detect modules and outputs bounding boxes along with label probabilities for objects within the image. The detect module consolidates the head's output, producing: bounding box coordinates for object localization, objectness scores to indicate the presence of objects, and class scores for identifying the detected object's category.

\subsubsection{Comparison with standard YOLOv11}
Table \ref{table1} compares the original YOLOv11 detectors with the proposed G-YOLOv11 detectors, focusing on the number of convolution filters in the Conv/GhostConv and C3k2/C3Ghost modules. In the G-YOLOv11 detectors, we halved the number of filters compared to the original YOLOv11 detectors. This reduction in the number of filters significantly decreases the computational complexity of the detector while maintaining a similar level of performance. For example, in the backbone of the G-YOLOv11 detectors, the number of filters in the Conv/GhostConv modules is halved (e.g., from [$16$, $32$, $64$, $128$, $256$] in YOLOv11n to [$8$, $16$, $32$, $64$, $128$] in G-YOLOv11n), and a similar reduction is seen in the C3Ghost modules. This reduction leads to a more lightweight architecture that is better suited for deployment in resource-constrained environments, such as CAD systems and mobile devices. Despite this reduction, the performance of G-YOLOv11 detectors remains comparable to their original counterparts, offering a favorable trade-off between efficiency and detection accuracy.

\begin{table}[h]
  \caption{Comparison of YOLOv11 and G-YOLOv11 detectors: number of convolution filters in Conv/GhostConv and C3k2/C3Ghost modules for backbone and head architectures.}
  \centering
  \begin{tabularx}{\textwidth}{l>{\centering\arraybackslash}p{6cm}>{\centering\arraybackslash}p{6cm}}
    \toprule
    Model & Backbone & Head \\
    \midrule
    YOLOv11n \cite{yolo11_ultralytics} & 
    \begin{tabular}{@{}c@{}}
        Conv: $5$ modules \\ 
        Filters: $[16, 32, 64, 128, 256]$ \\ 
        C3k2: $4$ modules \\ 
        Filters: $[64, 128, 128, 256]$ \\ 
    \end{tabular} & 
    \begin{tabular}{@{}c@{}}
        Conv: $2$ modules \\ 
        Filters: $[64, 128]$ \\ 
        C3k2: $4$ modules \\ 
        Filters: $[128, 64, 128, 256]$ \\
    \end{tabular} \\
    \hdashline
    \rowcolor{gray!30}
    \textbf{G-YOLOv11n (ours)} & 
    \begin{tabular}{@{}c@{}}
        GhostConv: $5$ modules \\ 
        Filters: $[8, 16, 32, 64, 128]$ \\ 
        C3Ghost: $4$ modules \\ 
        Filters: $[32, 64, 64, 128]$ \\ 
    \end{tabular} & 
    \begin{tabular}{@{}c@{}}
        GhostConv: $2$ modules \\ 
        Filters: $[32, 64]$ \\ 
        C3Ghost: $4$ modules \\ 
        Filters: $[64, 32, 64, 128]$ \\ 
    \end{tabular} \\
    \hdashline
    YOLOv11s \cite{yolo11_ultralytics} & 
    \begin{tabular}{@{}c@{}}
        Conv: $5$ modules \\ 
        Filters: $[32, 64, 128, 256, 512]$ \\ 
        C3k2: $4$ modules \\ 
        Filters: $[128, 256, 256, 512]$ \\ 
    \end{tabular} & 
    \begin{tabular}{@{}c@{}}
        Conv: $2$ layers \\ 
        Filters: $[128, 256]$ \\ 
        C3k2: $4$ modules \\ 
        Filters: $[256, 128, 256, 512]$ \\
    \end{tabular} \\
    \hdashline
    \rowcolor{gray!30}
    \textbf{G-YOLOv11s (ours)} & 
    \begin{tabular}{@{}c@{}}
        GhostConv: $5$ modules \\ 
        Filters: $[16, 32, 64, 128, 256]$ \\ 
        C3Ghost: $4$ modules \\ 
        Filters: $[64, 128, 128, 256]$ \\ 
    \end{tabular} & 
    \begin{tabular}{@{}c@{}}
        GhostConv: $2$ modules \\ 
        Filters: $[64, 128]$ \\ 
        C3Ghost: $4$ modules \\ 
        Filters: $[128, 64, 128, 256]$ \\ 
    \end{tabular} \\
    \hdashline
    YOLOv11m \cite{yolo11_ultralytics} & 
    \begin{tabular}{@{}c@{}}
        Conv: $5$ modules \\ 
        Filters: $[64, 128, 256, 512, 512]$ \\ 
        C3k2: $4$ modules \\ 
        Filters: $[256, 512, 512, 512]$ \\ 
    \end{tabular} & 
    \begin{tabular}{@{}c@{}}
        Conv: $2$ layers \\ 
        Filters: $[256, 512]$ \\ 
        C3k2: $4$ modules \\ 
        Filters: $[512, 256, 512, 512]$ \\
    \end{tabular} \\
    \hdashline
    \rowcolor{gray!30}
    \textbf{G-YOLOv11m (ours)} & 
    \begin{tabular}{@{}c@{}}
        GhostConv: $5$ modules \\ 
        Filters: $[32, 64, 128, 256, 512]$ \\ 
        C3Ghost: $4$ modules \\ 
        Filters: $[128, 256, 256, 512]$ \\ 
    \end{tabular} & 
    \begin{tabular}{@{}c@{}}
        GhostConv: $2$ modules \\ 
        Filters: $[128, 256]$ \\ 
        C3Ghost: $4$ modules \\ 
        Filters: $[256, 128, 256, 512]$ \\ 
    \end{tabular} \\
    \hdashline
    YOLOv11l \cite{yolo11_ultralytics} & 
    \begin{tabular}{@{}c@{}}
        Conv: $5$ modules \\ 
        Filters: $[64, 128, 256, 512, 512]$ \\ 
        C3k2: $8$ modules \\ 
        Filters: $[256, 512, 512, 512]$ \\ 
    \end{tabular} & 
    \begin{tabular}{@{}c@{}}
        Conv: $2$ layers \\ 
        Filters: $[256, 512]$ \\ 
        C3k2: $8$ modules \\ 
        Filters: $[512, 256, 512, 512]$ \\
    \end{tabular} \\
    \hdashline
    \rowcolor{gray!30}
    \textbf{G-YOLOv11l (ours)} & 
    \begin{tabular}{@{}c@{}}
        GhostConv: $5$ modules \\ 
        Filters: $[32, 64, 128, 256, 512]$ \\ 
        C3Ghost: $8$ modules \\ 
        Filters: $[128, 256, 256, 512]$ \\ 
    \end{tabular} & 
    \begin{tabular}{@{}c@{}}
        GhostConv: $2$ modules \\ 
        Filters: $[128, 256]$ \\ 
        C3Ghost: $8$ modules \\ 
        Filters: $[256, 128, 256, 512]$ \\ 
    \end{tabular} \\
    \hdashline
    YOLOv11x \cite{yolo11_ultralytics} & 
    \begin{tabular}{@{}c@{}}
        Conv: $5$ modules \\ 
        Filters: $[96, 192, 384, 768, 768]$ \\ 
        C3k2: $8$ modules \\ 
        Filters: $[384, 768, 768, 768]$ \\ 
    \end{tabular} & 
    \begin{tabular}{@{}c@{}}
        Conv: $2$ layers \\ 
        Filters: $[384, 768]$ \\ 
        C3k2: $8$ modules \\ 
        Filters: $[768, 384, 768, 768]$ \\
    \end{tabular} \\
    \hdashline
    \rowcolor{gray!30}
    \textbf{G-YOLOv11x (ours)} & 
    \begin{tabular}{@{}c@{}}
        GhostConv: $5$ modules \\ 
        Filters: $[48, 96, 192, 384, 768]$ \\ 
        C3Ghost: $8$ modules \\ 
        Filters: $[192, 384, 384, 768]$ \\ 
    \end{tabular} & 
    \begin{tabular}{@{}c@{}}
        GhostConv: $2$ modules \\ 
        Filters: $[192, 384]$ \\ 
        C3Ghost: $8$ modules \\ 
        Filters: $[384, 192, 384, 768]$ \\ 
    \end{tabular} \\
    \bottomrule
  \end{tabularx}
  \label{table1}
\end{table}

\subsection{Training details}
The detectors were initialized using MS COCO (Microsoft Common Objects in Context) pretrained weights and trained for $100$ epochs. The stochastic gradient descent optimizer was employed to optimize the detectors parameters, with a batch size set to $32$.

Implementation was based on the v8.3.55 release of Ultralytics YOLO, utilizing the PyTorch framework. All experiments were conducted on a cloud service equipped with an NVIDIA A10 24GB GPU.

\subsection{Evaluation metrics}
The detection performance of the detectors was evaluated using various metrics, including precision, recall, FScore, and mAP.

\textbf{Precision} measures the accuracy of the detector's predictions in terms of the proportion of true positives relative to all positive predictions. It is calculated as:
\begin{equation}
    Precision = \frac{TP}{TP+FP}
    \label{precision}
\end{equation}
where $TP$ and $FP$ represent true positives (correctly detected objects) and false positives (incorrectly detected objects), respectively.

\textbf{Recall} refers to the detector's ability to correctly detect all relevant objects present in the ground truth. It is computed as:
\begin{equation}
    Recall = \frac{TP}{TP+FN}
\end{equation}
where $FN$ is the number of objects present in the ground truth that the detector failed to detect.

\textbf{FScore} combines precision and recall into a single metric by calculating their harmonic mean. It is expressed as:
\begin{equation}
    FScore = \frac{2 \times Precision \times Recall}{Precision + Recall}
\end{equation}

\textbf{mAP} evaluates the detector’s ability to detect and localize objects across all classes. It is derived from the precision-recall (PR) curve, which summarizes the detector's performance across different thresholds for a given class. The AP for a single class is the area under the PR curve and is calculated as:
\begin{equation}
AP = \int_0^1 Precision(R) \, dR
\end{equation}
where $R$ represents recall, and $Precision(R)$ is the precision at a specific recall level.

mAP is the mean of the AP values across all classes. For $N$ classes, it is computed as:
\begin{equation}
   mAP = \frac{1}{N} \sum_{i=1}^{N} AP_i
\end{equation}
mAP is typically computed at an IoU threshold of $0.5$ (mAP@0.5) or across a range such as $0.5:0.95$ (mAP@0.5:0.95), with a commonly used step size of $0.05$. The IoU measures the overlap between the predicted bounding box and the ground truth, and mAP@0.5:0.95 offers a more stringent and robust evaluation.

\section{Results}
\label{section3}
\begin{figure}[H]
  \centering
  \includegraphics[height=8cm, width=\linewidth]{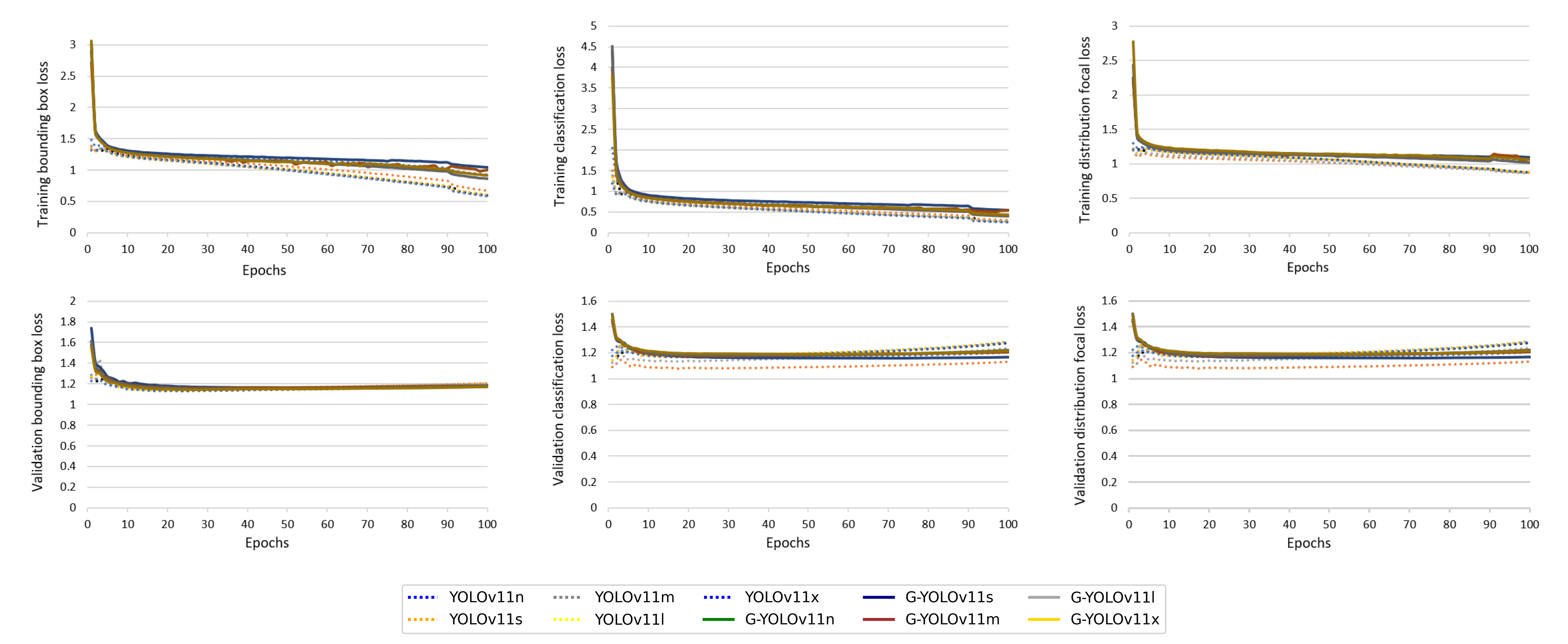}
  \caption{Plots of the training and validation loss for the YOLOv11 and proposed G-YOLOv11 detectors.}
  \label{figure6}
\end{figure}

\begin{figure}[h]
\centering
\begin{subfigure}{0.4956\textwidth}
    \centering 
    \includegraphics[height=6cm, width=\linewidth]{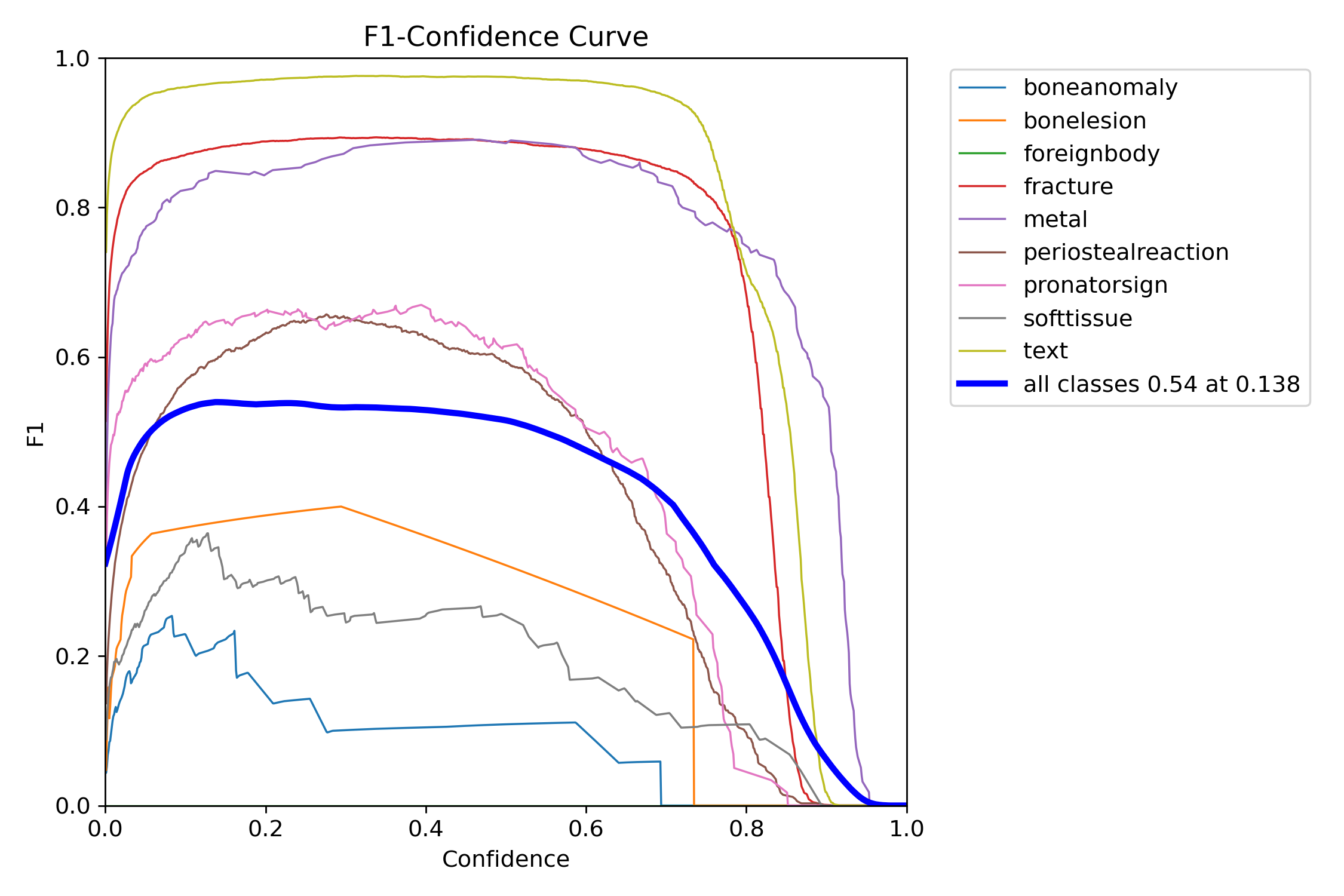}
\end{subfigure}
\hfill
\begin{subfigure}{0.4956\textwidth}
    \centering
    \includegraphics[height=6cm, width=\linewidth]{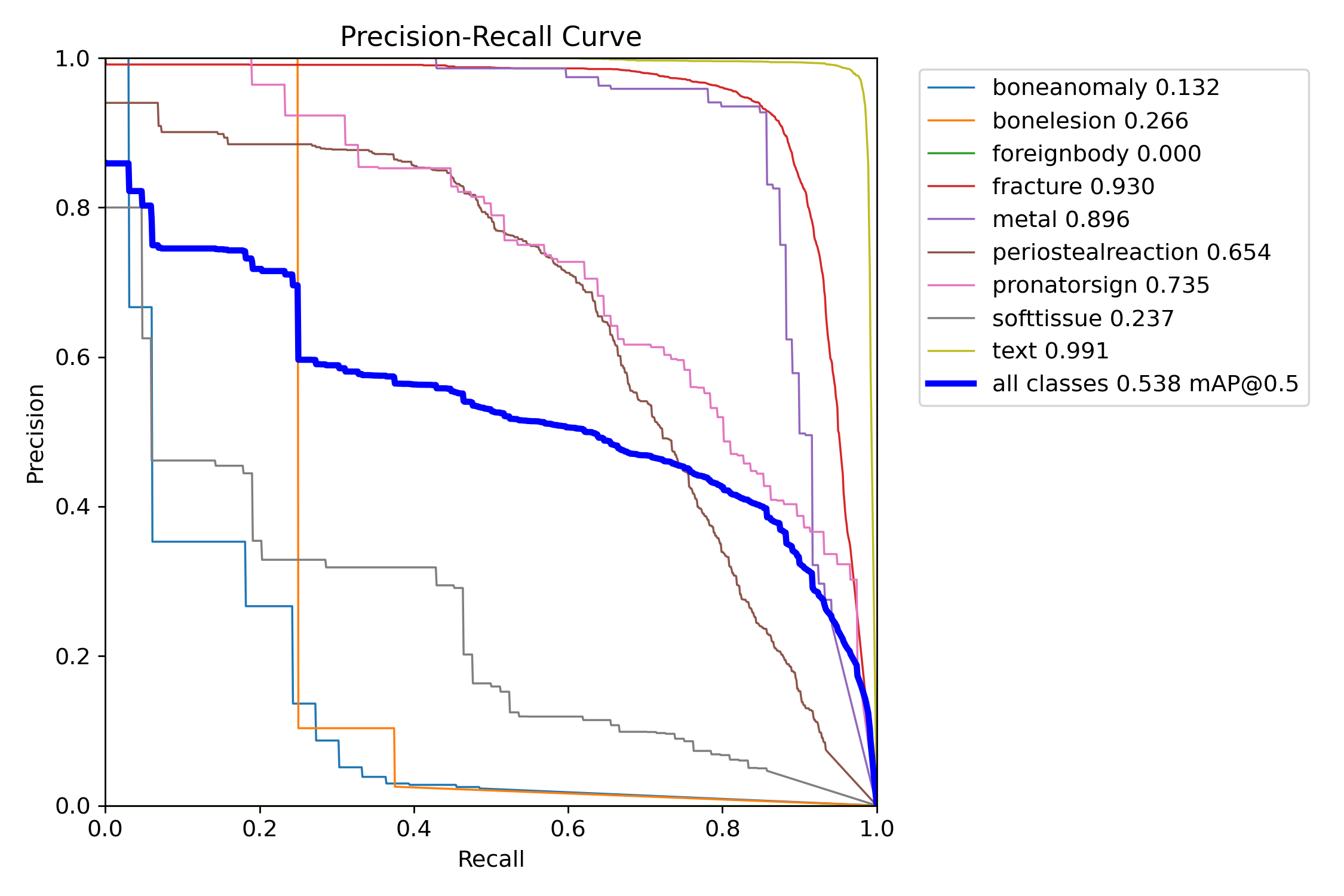}
\end{subfigure}
\caption{FScore as a function of confidence threshold, along with precision-recall curves, for the proposed G-YOLOv11l detector across all classes in the GRAZPEDWRI-DX validation set.}
\label{figure7}
\end{figure}

\begin{figure}[H]
    \centering
    \begin{subfigure}{0.19\textwidth}
        \includegraphics[width=3.2cm, height=3.5cm]{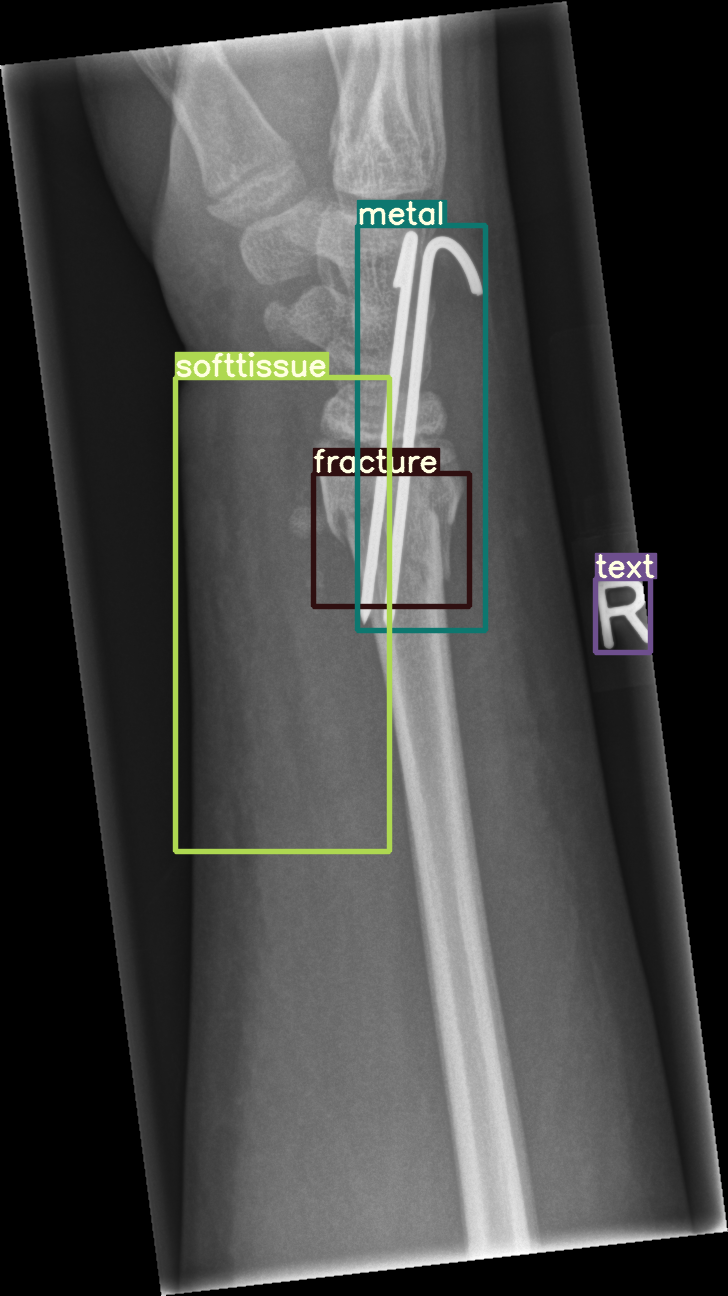}
        \caption{}
    \end{subfigure}
    \begin{subfigure}{0.19\textwidth}
        \includegraphics[width=3.2cm, height=3.5cm]{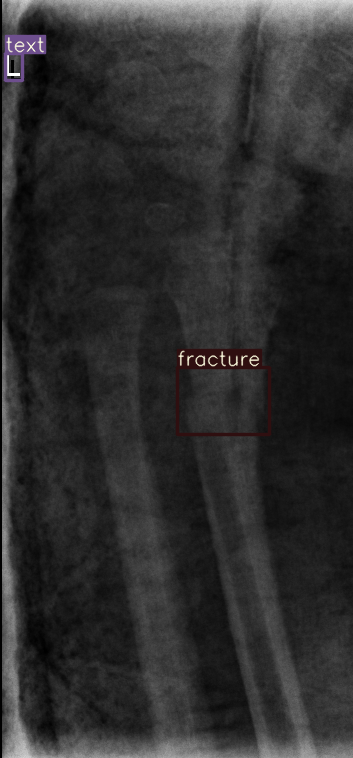}
        \caption{}
    \end{subfigure}
    \begin{subfigure}{0.19\textwidth}
        \includegraphics[width=3.2cm, height=3.5cm]{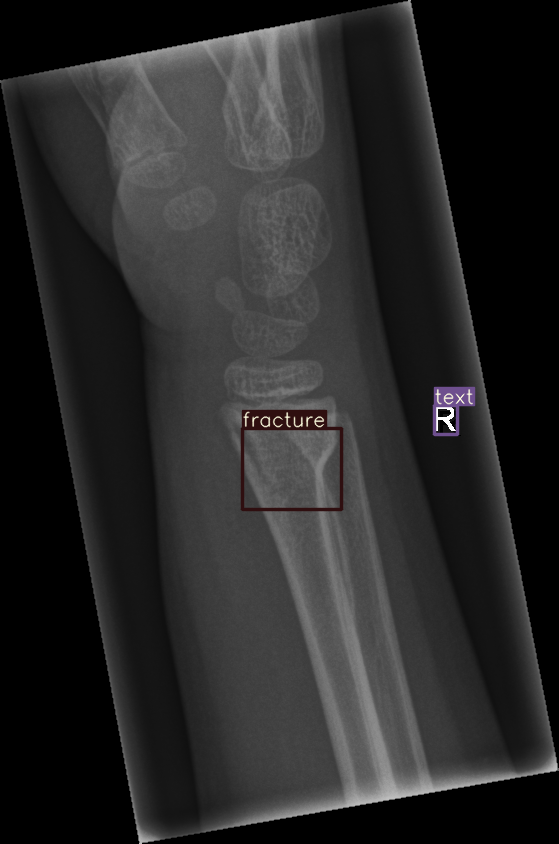}
        \caption{}
    \end{subfigure}
    \begin{subfigure}{0.19\textwidth}
        \includegraphics[width=3.2cm, height=3.5cm]{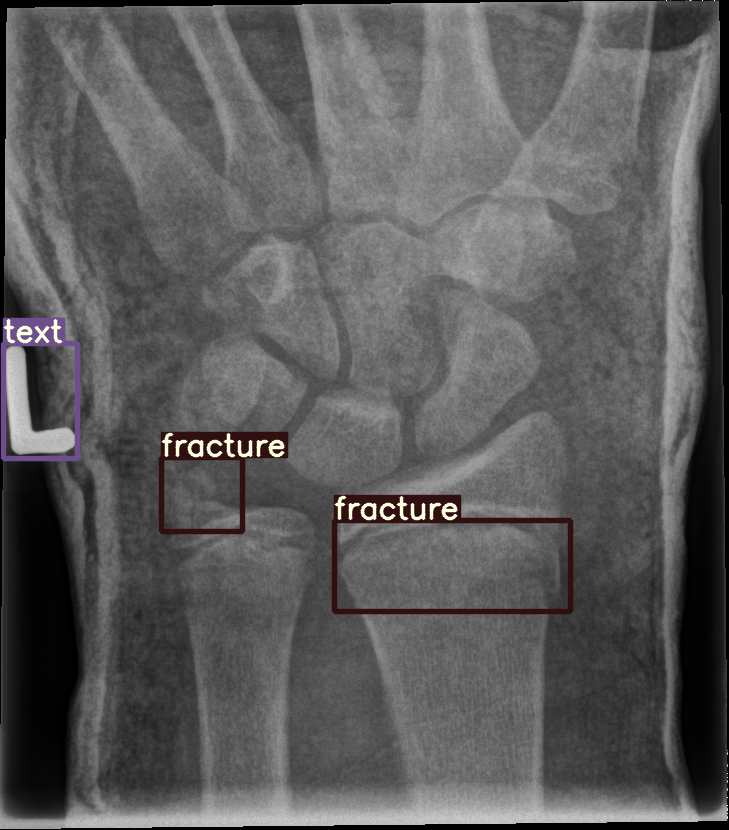}
        \caption{}
    \end{subfigure}
    \begin{subfigure}{0.19\textwidth}
        \includegraphics[width=3.2cm, height=3.5cm]{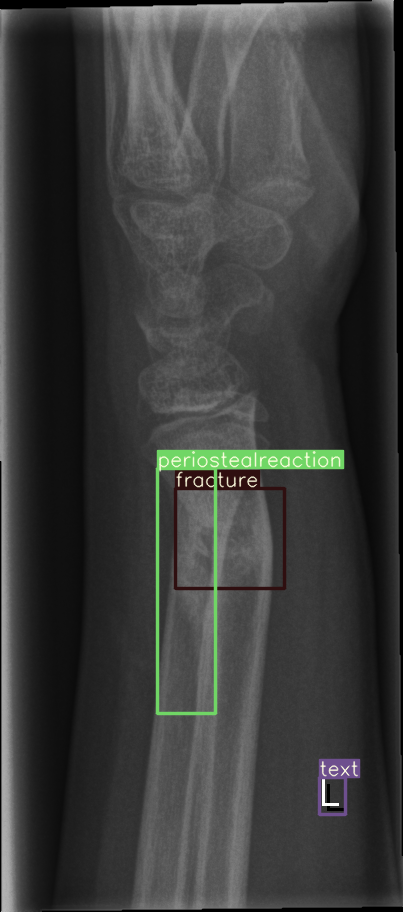}
        \caption{}
    \end{subfigure}

    \begin{subfigure}{0.19\textwidth}
        \includegraphics[width=3.2cm, height=3.5cm]{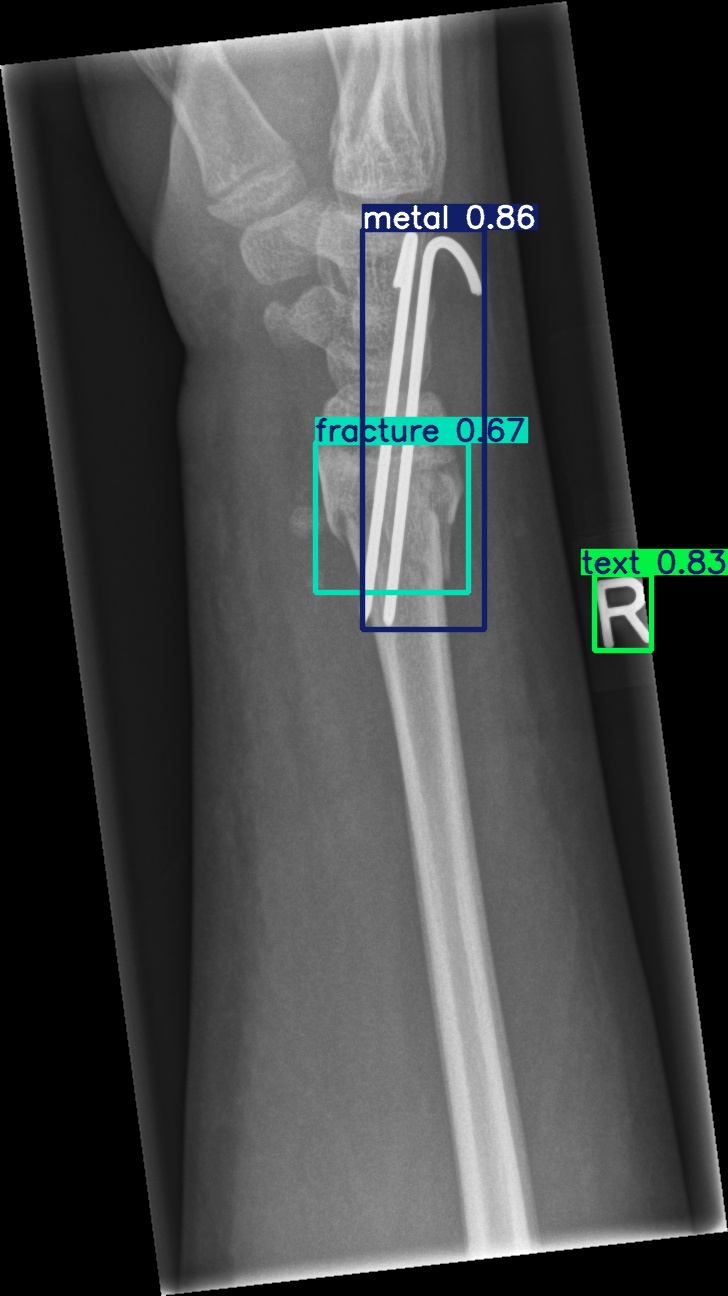}
        \caption{}
    \end{subfigure}
    \begin{subfigure}{0.19\textwidth}
        \includegraphics[width=3.2cm, height=3.5cm]{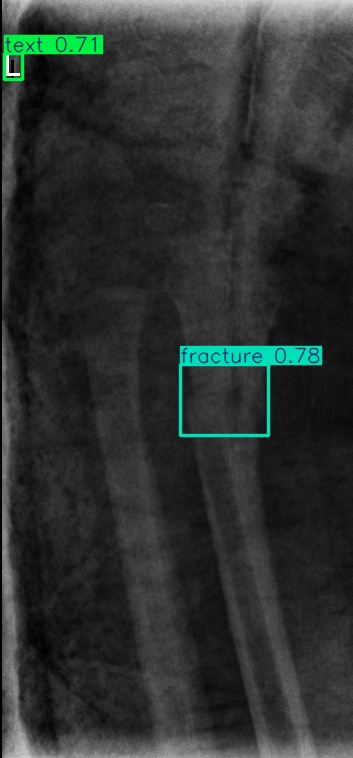}
        \caption{}
    \end{subfigure}
    \begin{subfigure}{0.19\textwidth}
        \includegraphics[width=3.2cm, height=3.5cm]{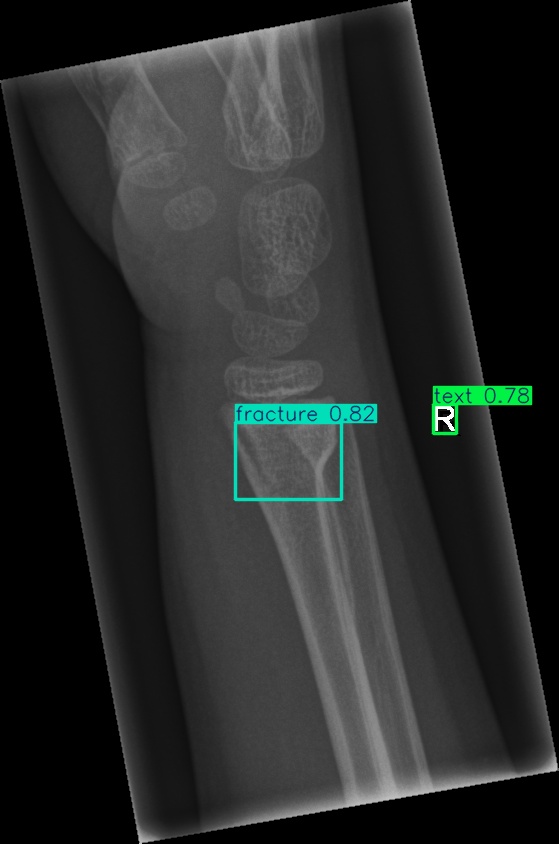}
        \caption{}
    \end{subfigure}
    \begin{subfigure}{0.19\textwidth}
        \includegraphics[width=3.2cm, height=3.5cm]{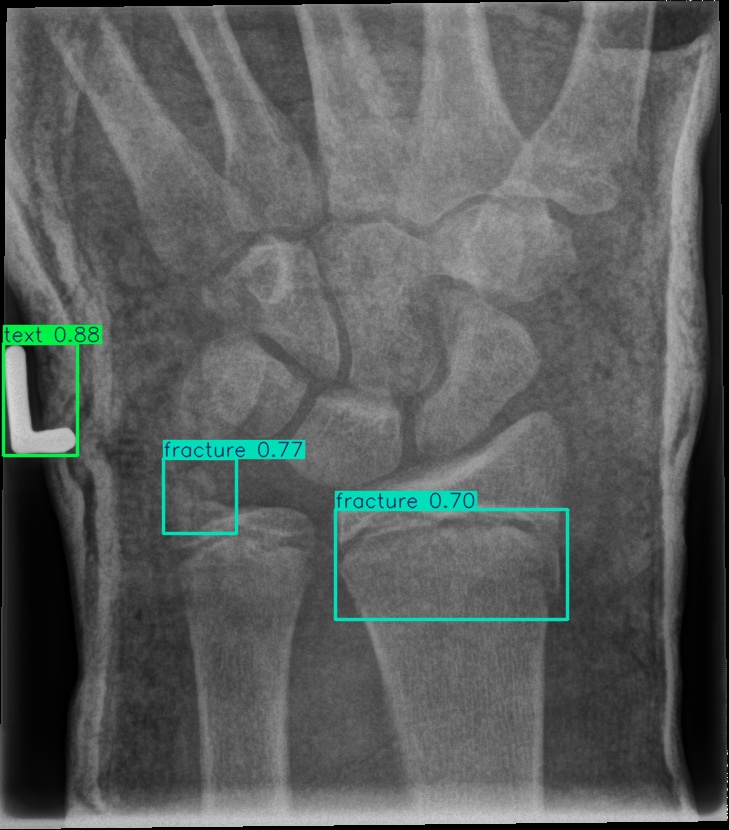}
        \caption{}
    \end{subfigure}
    \begin{subfigure}{0.19\textwidth}
        \includegraphics[width=3.2cm, height=3.5cm]{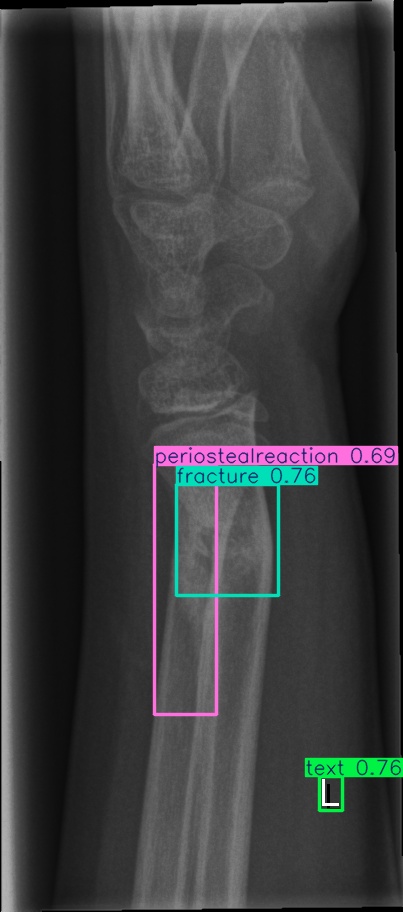}
        \caption{}
    \end{subfigure}

    \begin{subfigure}{0.19\textwidth}
        \includegraphics[width=3.2cm, height=3.5cm]{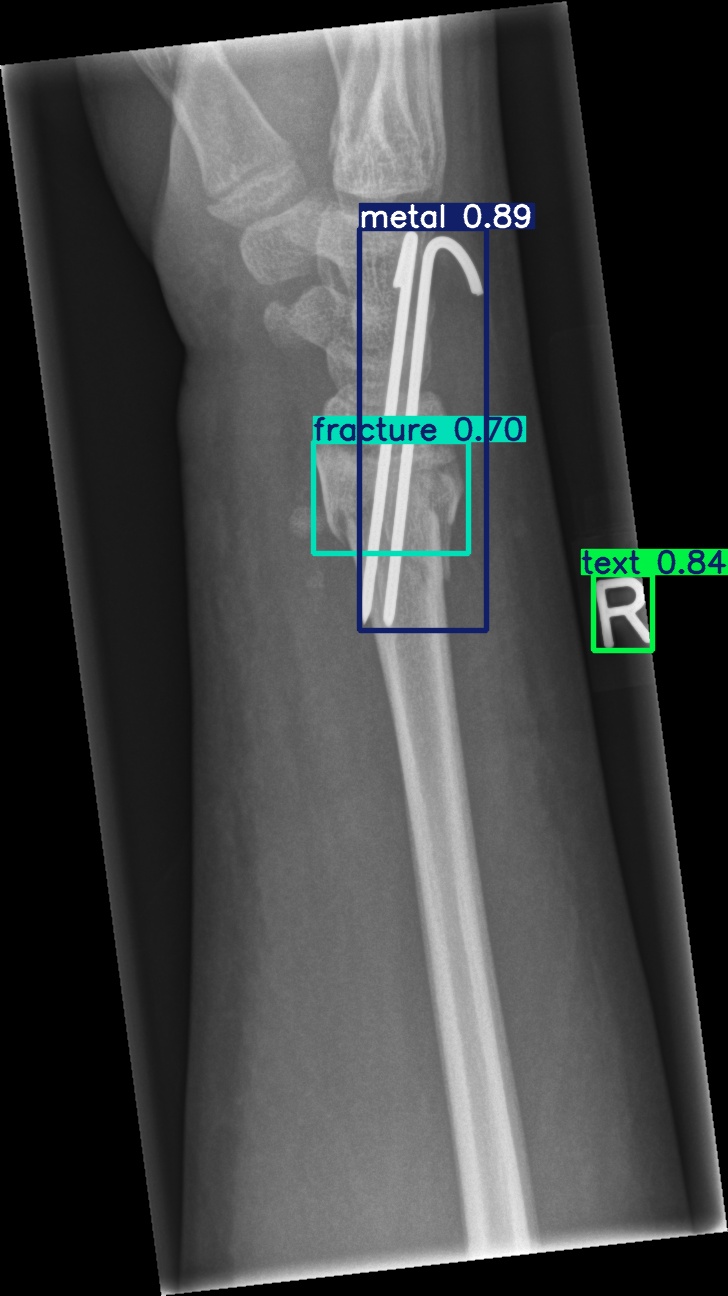}
        \caption{}
    \end{subfigure}
    \begin{subfigure}{0.19\textwidth}
        \includegraphics[width=3.2cm, height=3.5cm]{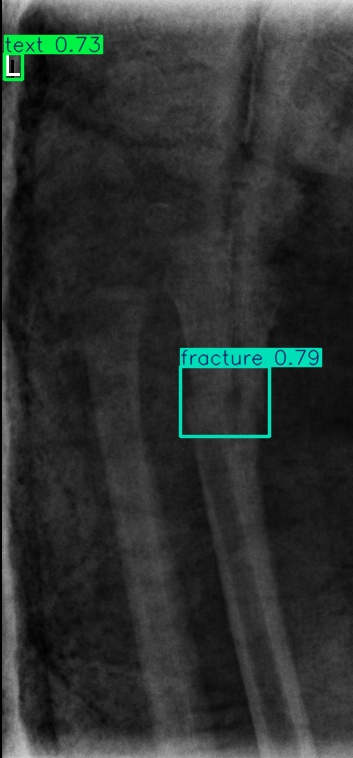}
        \caption{}
    \end{subfigure}
    \begin{subfigure}{0.19\textwidth}
        \includegraphics[width=3.2cm, height=3.5cm]{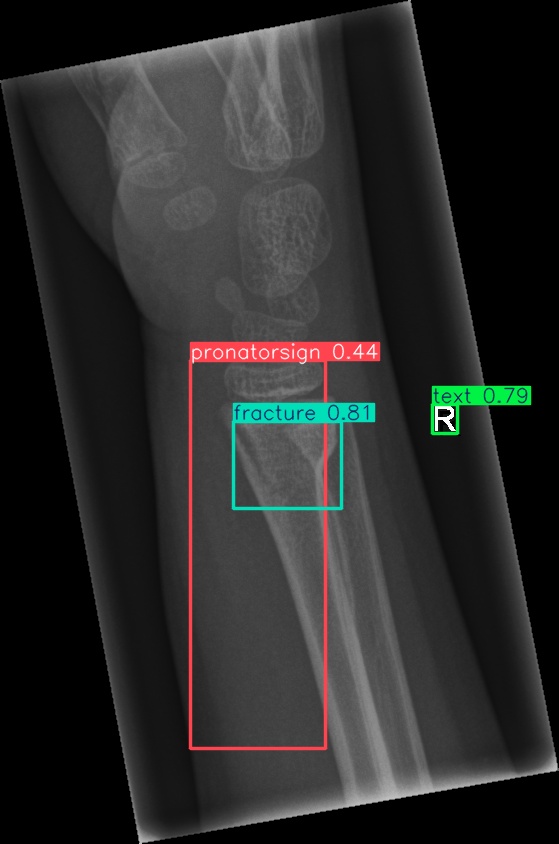}
        \caption{}
    \end{subfigure}
    \begin{subfigure}{0.19\textwidth}
        \includegraphics[width=3.2cm, height=3.5cm]{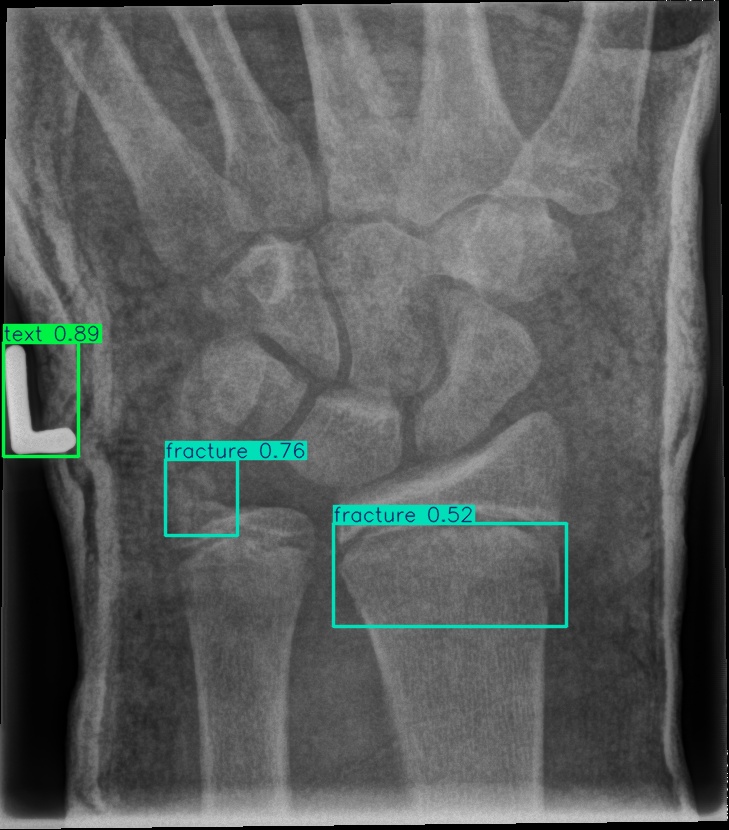}
        \caption{}
    \end{subfigure}
    \begin{subfigure}{0.19\textwidth}
        \includegraphics[width=3.2cm, height=3.5cm]{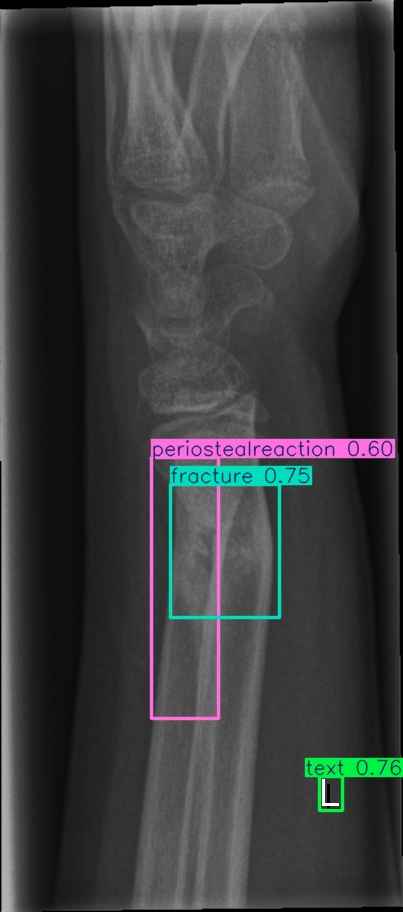}
        \caption{}
    \end{subfigure}
        
    \caption{Sample of five images from the GRAZPEDWRI-DX test set, displaying ground-truth and predicted bounding box annotations: (a-e): ground-truth annotations, (f-j): YOLOv11l annotations, and (k-o): G-YOLOv11 annotations.}
    \label{figure8}
\end{figure}

In this section, we first presented a comparison of the hardware requirements and the number of calculations for YOLOv11 and the proposed G-YOLOv11 in Table \ref{table2}, followed by their plots of training and validation loss in Figure \ref{figure6}. Next, we reported the performance of these detectors on the GRAZPEDWRI-DX test set in Table \ref{table3}. This is followed by qualitative and quantitative comparisons of detection using the YOLOv11l and proposed G-YOLOv11l detectors, as shown in Figure \ref{figure8} and Table \ref{table4}. Additionally, the FScore-confidence and PR curves are presented in Figure \ref{figure7}. Finally, we compared the proposed G-YOLOv11 with existing detectors in Table \ref{table5}. For a fair comparison, we reproduced the results of the existing detectors using the same dataset division employed in our experiments.

\begin{table}[H]
\caption{Comparison of hardware requirements, computational performance, and parameters for YOLOv11 and proposed G-YOLOv11 detectors on the GRAZPEDWRI-DX training set.}
\centering
\begin{tabular}{lccccccc}
\hline
Model & Layers & Params & Gradients & FLOPs & Detector size & GPU Memory & Train. time/ \\
& & (M) & (M) & (G) & (MB) & (GB) & epoch (Min) \\ 
\hline
YOLOv11n \cite{yolo11_ultralytics} & $\mathbf{319}$ & $2.591$ & $2.591$ & $6.4$ & $5.2$ & $4.77$ & $02.19$ \\
\rowcolor{gray!30}
\textbf{G-YOLOv11n (ours)} & $397$ & $\mathbf{0.676}$ & $\mathbf{0.675}$ & $\mathbf{2.3}$ & $\mathbf{1.6}$ & $\mathbf{3.64}$ & $\mathbf{02.11}$\\
YOLOv11s \cite{yolo11_ultralytics} & $\mathbf{319}$ & $9.431$ & $9.431$ & $21.6$ & $18.3$ & $8.21$ & $3.34$ \\
\rowcolor{gray!30}
\textbf{G-YOLOv11s (ours)} & $397$ & $\mathbf{2.039}$ & $\mathbf{2.039}$ & $\mathbf{6.1}$ & $\mathbf{4.2}$ & $\mathbf{5.14}$ & $\mathbf{02.53}$\\
YOLOv11m \cite{yolo11_ultralytics} & $\mathbf{409}$ & $20.095$ & $20.059$ & $68.2$ & $38.6$ & $16.5$ & $7.08$\\
\rowcolor{gray!30}
\textbf{G-YOLOv11m (ours)} & $397$ & $\mathbf{7.202}$ & $\mathbf{7.202}$ & $\mathbf{20.1}$ & $\mathbf{14}$ & $\mathbf{8.88}$ & $\mathbf{04.19}$ \\
YOLOv11l \cite{yolo11_ultralytics} & $\mathbf{631}$ & $25.317$ & $25.317$ & $87.3$ & $48.8$ & $20.8$ & $9.20$\\
\rowcolor{gray!30}
\textbf{G-YOLOv11l (ours)} & $579$ & $\mathbf{7.794}$ & $\mathbf{7.794}$ & $\mathbf{21.0}$ & $\mathbf{15.3}$ & $\mathbf{10.2}$ & $\mathbf{05.04}$\\
YOLOv11x \cite{yolo11_ultralytics} & $\mathbf{631}$ & $56.884$ & $56.884$ & $195.5$ & $109.1$ & $8.39$ & $17$\\
\rowcolor{gray!30}
\textbf{G-YOLOv11x (ours)} & $579$ & $\mathbf{16.920}$ & $\mathbf{16.920}$ & $\mathbf{44.5}$ & $\mathbf{32.7}$ & $\mathbf{4.09}$ & $\mathbf{08.39}$\\
\hline
\end{tabular}
\label{table2}
\end{table}

\begin{table}[h]
  \caption{Performance comparison of YOLOv11 and Proposed G-YOLOv11 detectors on the GRAZPEDWRI-DX test set. Speed represents the total time for preprocessing, inference, and post-processing a single image.}
  \centering
    \begin{tabular}{lccccc}
    \toprule
    Model & Precision & Recall & mAP@0.5 & mAP@0.5:0.95 & Speed (ms) \\ 
    \hline
    YOLOv11n \cite{yolo11_ultralytics} & $\mathbf{0.812}$ & $\mathbf{0.509}$ & $\mathbf{0.545}$ & $\mathbf{0.339}$ & $1.8$ \\
    \rowcolor{gray!30}
    \textbf{G-YOLOv11n (ours)} & $0.769$ & $0.5$ & $0.522$ & $0.319$ & $\mathbf{1.6}$ \\ 
    YOLOv11s \cite{yolo11_ultralytics} & $\mathbf{0.79}$ & $\mathbf{0.55}$ & $\mathbf{0.554}$ & $\mathbf{0.351}$ & $2.4$ \\
    \rowcolor{gray!30}
    \textbf{G-YOLOv11s (ours)} & $0.784$ & $0.54$ & $0.545$ & $0.334$ & $\mathbf{1.9}$ \\ 
    YOLOv11m \cite{yolo11_ultralytics} & $\mathbf{0.827}$ & $\mathbf{0.537}$ & $\mathbf{0.577}$ & $\mathbf{0.361}$ & $4.9$ \\
    \rowcolor{gray!30}
    \textbf{G-YOLOv11m (ours)} & $0.768$ & $0.534$ & $0.541$ & $0.339$ & $\mathbf{2.8}$\\
    YOLOv11l \cite{yolo11_ultralytics} & $0.646$ & $\mathbf{0.702}$ & $\mathbf{0.619}$ & $\mathbf{0.385}$ & $6$ \\
    \rowcolor{gray!30}
    \textbf{G-YOLOv11l (ours)} & $\mathbf{0.806}$ & $0.493$ & $0.535$ & $0.341$ & $\mathbf{3}$\\
    YOLOv11x \cite{yolo11_ultralytics} & $\mathbf{0.766}$ & $\mathbf{0.571}$ & $\mathbf{0.614}$ & $\mathbf{0.381}$ & $10$\\
    \rowcolor{gray!30}
    \textbf{G-YOLOv11x (ours)} & $0.67$ & $0.548$ & $0.562$ & $0.352$ & $\mathbf{4.5}$\\ 
    \bottomrule
  \end{tabular}
  \label{table3}
\end{table}

\begin{table}[H]
\caption{Confidence scores predicted by YOLOv11l and the proposed G-YOLOv11l detectors for five images from the GRAZPEDWRI-DX test set, PR: pronator sign, FP: False positive.}
\centering
\begin{tabular}{lccc}
\hline
Figure number & Class name & Confidence score predicted & Confidence score predicted\\
& & by YOLOv11l & by our G-YOLOv11l\\
\hline
8(a) & soft tissue, fracture, metal, text & -, $0.67$, $0.86$, $0.83$ & -, $0.70$, $0.89$, $0.84$ \\
8(b) & fracture, text & $0.78$, $0.71$ & $0.79$, $0.73$ \\
8(c) & fracture, text & $0.82$, $0.78$ & $0.81$, $0.79$, (FP: pr $0.44)$ \\
8(d) & fracture, fracture, text & $0.77$, $0.70$, $0.88$ & $0.76$, $0.52$, $0.89$ \\
8(e) & fracture, periosteal reaction, text & $0.76$, $0.69$, $0.76$ & $0.75$, $0.60$, $0.76$ \\
\hline
\end{tabular}
\label{table4}
\end{table}

\begin{table}[t]
  \caption{Detection results of existing detectors and the proposed G-YOLOv11 on the GRAZPEDWRI-DX test set. All detectors are in large configuration. Inference time is measured for a single image on an NVIDIA A10 GPU.}
  \centering
    \begin{tabular}{lcccccc}
    \toprule
    Model & FScore & mAP@0.5 & mAP@0.5:0.95 & Params (M) & FLOPs (G) & Inference (ms)\\ 
    \hline
    YOLOv8 \cite{ju2023fracture} & $0.698$ & $\mathbf{0.672}$ & $0.403$ & $43.636$ & $165.4$ & $6.2$\\
    YOLOv9-C \cite{chien2024yolov9} & $0.682$ & $0.617$ & $0.389$ & $25.536$ & $103.7$ & $5.8$\\
    YOLOv9-E \cite{chien2024yolov9} & $\mathbf{0.711}$ & $0.657$ & $\mathbf{0.442}$ & $58.152$ & $192.7$ & $13.0$ \\ 
    YOLOv10 \cite{ahmed2024pediatric} & $0.604$ & $0.568$ & $0.36$ & $25.779$ & $127.3$ & $6.7$ \\ 
    YOLOv11 \cite{das5056626detection} & $0.673$ & $0.619$ & $0.385$ & $25.317$ & $87.3$ & $5.5$\\
    \rowcolor{gray!30}
    \textbf{G-YOLOv11 (ours)} & $0.612$ & $0.535$ & $0.341$ & $\mathbf{7.794}$ & $\mathbf{21.0}$ & $\mathbf{2.4}$\\ 
    \bottomrule
  \end{tabular}
  \label{table5}
\end{table}

\section{Discussion}
\label{section4}
In this section, we first presented a qualitative and quantitative comparison of the performance of the proposed G-YOLOv11 with YOLOv11. Subsequently, we discussed the results of the proposed G-YOLOv11 in comparison to existing detectors.

\subsection{Qualitative Comparison with Standard YOLOv11}
Figure \ref{figure6} compares the training and validation losses of YOLOv11 and G-YOLOv11 models over $100$ epochs. Both models show consistent loss reductions, with G-YOLOv11 achieving slightly lower losses, indicating better optimization and generalization. Smaller variants (e.g., YOLOv11n, YOLOv11s) have higher losses than larger ones (e.g., YOLOv11l, YOLOv11x), reflecting performance-complexity trade-offs. 

Figure \ref{figure7} shows the FScore-Confidence and PR curves for the G-YOLOv11 detector on the GRAZPEDWRI-DX validation set. The FScore-Confidence curve facilitates the identification of the confidence threshold at which the model achieves the best FSCore (a balance between precision and recall) for each class. This curve indicates that classes like "fracture" and "text" achieve high FScores across confidence thresholds, while others, such as "bone anaomaly," show variability. The PR curve highlights strong performance for "fracture" and "text", with high precision and recall, whereas classes like "bone anomaly" experience steep precision declines at higher recall. These findings underscore the class imbalance in the GRAZPEDWRI-DX training set (Figure \ref{figure3}), which impacts the G-YOLOv11's ability to effectively learn classes with fewer instances.

The results in Table \ref{table4}, derived from Figure \ref{figure8}, compares the confidence scores predicted by YOLOv11l and G-YOLOv11l detectors for five test images from the GRAZPEDWRI-DX dataset. Figure \ref{figure8} contrasts ground-truth annotations with predictions from YOLOv11l and G-YOLOv11l. G-YOLOv11l shows slight improvements, such as higher confidence scores for "fracture" and "metal" in Figure \ref{figure8}(a) ($0.70$ and $0.89$ vs. $0.67$ and $0.86$) and for "text" in Figure \ref{figure8}(b) ($0.73$ vs. $0.71$). However, limitations include false positives, such as for "pronator sign" in Figure \ref{figure8}(c) (score $0.44$), and lower confidence for some instances, as seen in Figures \ref{figure8}(d) and \ref{figure8}(e). While G-YOLOv11l demonstrates competitive performance, further refinements are required to improve its robustness.

\subsection{Quantitative Comparison with Standard YOLOv11}
The efficiency metrics from Tables \ref{table2} and \ref{table3} demonstrate that G-YOLOv11 offers substantial improvements over YOLOv11 across all configurations in terms of computational resource requirements. First, the proposed G-YOLOv11 reduces detector complexity, with fewer parameters and gradients. For example, G-YOLOv11m has $7.202$ million parameters, a $64.2\%$ reduction compared to YOLOv11m’s $20.095$ million, while FLOPs for G-YOLOv11x decrease by $77.2\%$ from $195.5$ GFLOPs to $44.5$ GFLOPs. Second, G-YOLOv11 requires less memory and storage, with models consistently smaller in size and GPU memory consumption. For example, G-YOLOv11l is $15.3$ MB compared to YOLOv11l’s $48.8$ MB, a $68.7\%$ reduction, and G-YOLOv11x uses $4.09$ GB of memory versus YOLOv11x’s $8.39$ GB. Third, training efficiency is improved, with G-YOLOv11 models requiring shorter training times per epoch, such as G-YOLOv11m reducing training time from $7.08$ to $4.19$ minutes per epoch, a $40.8\%$ improvement. Finally, inference speed is comparable or faster, with similar speeds for nano models and faster inference for larger models. These results highlight the efficiency and scalability of G-YOLOv11.

In terms of accuracy, Table \ref{table3} shows that the YOLOv11 detectors consistently slightly outperform the proposed G-YOLOv11 detectors across most evaluation metrics by at most $8.4\%$ in mAP@0.5 and $4.4\%$ in mAP@0.5:0.95. This suggests that the proposed G-YOLOv11 detector prioritizes efficiency over detection performance, resulting in slight sacrifices in accuracy metrics. However, the differences in accuracy are not drastic, indicating that G-YOLOv11 detectors maintain competitive performance despite being more computationally efficient.

From this analysis, despite slight accuracy reductions, the proposed G-YOLOv11 demonstrates competitive performance while achieving significant computational savings, making it a compelling alternative for scenarios where efficiency is paramount, such as CAD systems.

\subsection{Comparison with State-of-the-art}
The results presented in Table \ref{table5} compare the detection performance of several existing detectors with the proposed G-YOLOv11 on the GRAZPEDWRI-DX test set. While G-YOLOv11 achieves the lowest mAP metrics ($0.535$ and $0.341$), it significantly outperforms other detectors in computational efficiency. With only $7.794$ million parameters, $21.0$ GFLOPs, and a $2.4$ ms inference time, G-YOLOv11 is the most lightweight and fastest model, making it particularly well-suited for resource-constrained environments and real-time applications. In contrast, YOLOv9-E delivers the best detection performance (FScore $0.711$, mAP@0.5:0.95 $0.442$), but incurs a substantially higher computational cost. 

In conclusion, although G-YOLOv11 compromises detection performance to some extent, its exceptional efficiency makes it a compelling choice for scenarios where real-time processing and limited computational resources are critical. Further refinements could enhance its detection accuracy without diminishing its efficiency.

\section{Conclusion and Future Work}
\label{section5}

We introduced a novel YOLO-based CAD system for detecting pediatric wrist trauma in X-ray images. The system leverages our G-YOLOv11 detector, a lightweight adaptation of YOLOv11 incorporating ghost convolutions, which significantly reduces computational resource requirements and facilitates deployment in clinical settings. The proposed G-YOLOv11l detector achieved an mAP@0.5 of $0.535$ with an inference time of $2.4$ ms on an NVIDIA A10 GPU, setting new state-of-the-art benchmark in efficiency and outperforming existing detectors. Future work will investigate the impact of integrating ghost convolutions into other YOLO versions to enhance both efficiency and accuracy. Additionally, we aim to develop innovative and efficient convolutional operations to further advance the field of fracture detection.

\bibliographystyle{unsrt}  
\bibliography{references}

\end{document}